\documentclass[fleqn,twoside]{article}
\usepackage{espcrc2}

\usepackage{graphicx}
\usepackage[figuresright]{rotating}

\newcommand{\AmS}{{\protect\the\textfont2
  A\kern-.1667em\lower.5ex\hbox{M}\kern-.125emS}}

\title{Lepton flavour universality test at the CERN NA62 experiment}

\author{Evgueni Goudzovski\address[MCSD]{School of Physics and Astronomy,
        University of Birmingham,\\
        Edgbaston, Birmingham, B15 2TT, United Kingdom}
        on behalf of the NA62 collaboration}

\begin{document}

\begin{abstract}
A precision test of lepton universality by measurement of the ratio
$R_K$ of $K^+\to e^+\nu$ to $K^+\to\mu^+\nu$ decay rates was
performed using a sample of 59963 $K^+\to e^+\nu$ candidates with
8.8\% background contamination collected by the CERN NA62
experiment. The result $R_K=(2.486\pm0.013)\times10^{-5}$ is in
agreement with the Standard Model expectation.\vspace{1pc}
\end{abstract}

\maketitle

\section{Introduction}

Decays of pseudoscalar mesons to light leptons are suppressed in the
Standard Model (SM) by angular momentum conservation. In particular,
the SM width of $P^\pm\to l^\pm\nu$ decays (with
$P=\pi,K,D_{(s)},B$) is
\begin{displaymath}
\Gamma^\mathrm{SM} = \frac{G_F^2 M_P M_\ell^2}{8\pi}
\left(1-\frac{M_\ell^2}{M_P^2}\right)^2 f_P^2|V_{qq\prime}|^2,
\end{displaymath}
where $G_F$ is the Fermi constant, $M_P$ and $M_\ell$ are meson and
lepton masses, $f_P$ is the decay constant, and $V_{qq\prime}$ is
the corresponding CKM matrix element.

Within the two Higgs doublet models (2HDM of type II), including the
minimal supersymmetric one, the charged Higgs boson ($H^\pm$)
exchange induces a tree-level contribution to (semi)leptonic decays
proportional to the Yukawa couplings of quarks and
leptons~\cite{ho93}. In $P^\pm\to\ell^\pm\nu$, it can compete with
the $W^\pm$ exchange due to the helicity suppression of the latter.
At tree level, the $H^\pm$ exchange contribution is lepton flavour
independent, and for $P=\pi, K, B$ leads to~\cite{is06}
\begin{displaymath}
\frac{\Gamma(P^\pm\to\ell^\pm\nu)}{\Gamma^{\rm
SM}(P^\pm\to\ell^\pm\nu)}\!=\!\left[1-\left(\frac{M_P}{M_H}\right)^2
\frac{\tan^2\beta}{1+\varepsilon_0\tan\beta}\right]^2.
\end{displaymath}
Here $M_H$ is the $H^\pm$ mass, $\tan\beta$ is the ratio of the two
Higgs vacuum expectation values, and $\varepsilon_0\approx 10^{-2}$
is an effective coupling.

A plausible choice of parameters $M_H=500~{\rm GeV}/c^2$,
$\tan\beta=40$ leads to $\sim 30\%$ relative suppression of
$B^+\to\ell^+\nu$ decays, and $\sim 0.3\%$ suppression of
$K^+,D_s^+\to\ell^+\nu$ decays with respect to their SM rates.
However, searches for new physics in the decay rates are hindered by
the uncertainties of their SM predictions. In particular,
interpretation of the measurements of the ratio
$\Gamma(K^+\to\mu^+\nu)/\Gamma(K^+\to\pi^0\mu^+\nu)$ in terms of
constraints on ($M_H$, $\tan\beta$) phase space is currently limited
by lattice QCD uncertainties~\cite{an10}.

On the other hand, the ratio of kaon leptonic decay widths
$R_K=\Gamma(K_{e2})/\Gamma(K_{\mu 2})$, where the notation $K_{\ell
2}$ is adopted for $K^+\to\ell^+\nu$ decays, is sensitive to
loop-induced lepton flavour violating (LFV) effects via the $H^\pm$
exchange~\cite{ma06}:
\begin{displaymath}
\Delta R_K/R_K^\mathrm{SM} \simeq \left(\frac{M_K}{M_H}\right)^4
\left(\frac{M_\tau}{M_e}\right)^2 |\Delta _R^{31}|^2\tan^6\beta,
\end{displaymath}
where the mixing parameter between the superpartners of the
right-handed leptons $|\Delta_{R}^{31}|$ can reach $\sim 10^{-3}$.
This can enhance $R_K$ by ${\cal O}(1\%)$ relative without
contradicting any presently known experimental constraints,
including upper bounds on the LFV decays $\tau\to eX$ with
$X=\eta,\gamma,\mu\mu$.

Unlike the individual $K_{\ell 2}$ decay widths, the ratio
$R_K=\Gamma(K_{e2})/\Gamma(K_{\mu 2})$ is precisely predicted within
the SM due to cancellation of hadronic uncertainties~\cite{ci07}:
\begin{eqnarray*}
\label{Rdef}
R_K^\mathrm{SM}\!\!&\!=\!\!&\left(\frac{M_e}{M_\mu}\right)^2
\left(\frac{M_K^2-M_e^2}{M_K^2-M_\mu^2}\right)^2 (1 + \delta
R_{\mathrm{QED}})=\\
&\!=\!\!&(2.477 \pm 0.001)\times 10^{-5},
\end{eqnarray*}
where $\delta R_{\mathrm{QED}}=(-3.79\pm0.04)\%$ is an
electromagnetic correction due to the internal bremsstrahlung (IB)
process.

The sensitivity to LFV and the precision of the SM prediction make
$R_K$ an excellent probe of lepton universality. The current world
average (based on final results only) $R_K^{\rm
WA}=(2.490\pm0.030)\times 10^{-5}$ is dominated by a recent KLOE
result~\cite{am09}. A precise measurement of $R_K$ based on a part
(40\%) of the data sample collected by the CERN NA62 experiment in
2007 is reported here. This is an update of an earlier result
obtained with the same data sample~\cite{go10}.

\section{Beam, detector and data taking}

The NA48/2 experimental setup~\cite{fa07} has been used for the NA62
2007--08 data taking. Experimental conditions have been optimized
for the $K_{e2}/K_{\mu2}$ measurement. The beam line is designed to
deliver simultaneous unseparated $K^+$ and $K^-$ beams derived from
the SPS 400 GeV/$c$ primary protons. However, the muon sweeping
system was optimized for the positive beam in 2007, and the sample
used for the present analysis was collected with the $K^+$ beam
only. Positively charged particles within a narrow momentum band of
$(74.0\pm1.6)$ GeV/$c$ are selected by an achromatic system of four
dipole magnets with zero total deflection, pass through a muon
sweeping system, and enter a fiducial decay volume contained in a
114 m long cylindrical vacuum tank.

With about $1.8\times 10^{12}$ primary protons incident on the
target per SPS pulse of about $4.8$~s duration, the secondary beam
flux at the entrance to the decay volume is $2.5\times 10^7$
particles per pulse, of which 5\% are kaons ($K^+$). The fraction of
beam kaons decaying in the vacuum tank at nominal momentum is
$18\%$. The transverse size of the beam within the decay volume is
$\delta x = \delta y = 7$~mm (rms), and its angular divergence is
negligible.

Among the subdetectors located downstream the decay volume, a
magnetic spectrometer, a plastic scintillator hodoscope (HOD) and a
liquid krypton electromagnetic calorimeter (LKr) are principal for
the measurement. The spectrometer, used to detect charged products
of kaon decays, is composed of four drift chambers (DCHs) and a
dipole magnet. The HOD producing fast trigger signals consists of
two planes of strip-shaped counters. The LKr, used for particle
identification and as a veto, is an almost homogeneous ionization
chamber, $27X_{0}$ deep, segmented transversally into 13,248 cells
($2\times 2~{\rm cm}^2$ each), and with no longitudinal
segmentation. A beam pipe traversing the centres of the detectors
allows undecayed beam particles and muons from decays of beam pions
to continue their path in vacuum.

A minimum bias trigger configuration has been employed. The $K_{e2}$
trigger condition consists of coincidence of hits in the two HOD
planes (the $Q_{1}$ signal), loose lower and upper limits on DCH hit
multiplicity (the 1-track signal), and LKr energy deposit
$(E_\mathrm{LKr})$ of at least 10 GeV. The $K_{\mu2}$ trigger
condition requires a coincidence of the $Q_1$ and 1-track signals
downscaled by a factor $D=150$.

\section{Measurement strategy, event selection}

The analysis strategy is based on counting the numbers of
reconstructed $K_{e2}$ and $K_{\mu 2}$ candidates collected
concurrently. Thus the analysis does not rely on the absolute beam
flux measurement, and several systematic effects (e.g. due to
charged track reconstruction and $Q_1$ trigger efficiencies,
time-dependent effects) cancel at first order.

Due to the significant acceptance and background dependence, the
measurement is performed independently in 10 bins of lepton momentum
covering a range from 13 to 65~GeV/$c$. The first bin spans 7
GeV/$c$, while the others are 5 GeV/$c$ wide. The selection
conditions have been optimized separately in each momentum bin. The
ratio $R_K$ in each bin is computed as
\begin{eqnarray*}
R_K &=& \frac{1}{D}\cdot \frac{N(K_{e2})-N_{\rm B}(K_{e2})}{N(K_{\mu
2}) - N_{\rm B}(K_{\mu 2})}\cdot \frac{A(K_{\mu 2})}{A(K_{e2})}\times\\
&&\frac{f_\mu\times\epsilon(K_{\mu 2})}
{f_e\times\epsilon(K_{e2})}\cdot\frac{1}{f_\mathrm{LKr}},
\end{eqnarray*}
where $N(K_{\ell 2})$ are the numbers of selected $K_{\ell 2}$
candidates $(\ell=e,\mu)$, $N_{\rm B}(K_{\ell 2})$ are the numbers
of background events, $A(K_{\mu 2})/A(K_{e2})$ is the geometric
acceptance correction, $f_\ell$ are the lepton identification
efficiencies, $\epsilon(K_{\ell 2})$ are the trigger efficiencies,
$f_\mathrm{LKr}$ is the global efficiency of the LKr readout, and
$D=150$ is the $K_{\mu 2}$ trigger downscaling factor.

To evaluate the acceptance correction and the geometric parts of the
acceptances for background processes, a detailed Monte Carlo (MC)
simulation including beam line optics, full detector geometry and
material description, magnetic fields, local inefficiencies of DCH
wires, non-working LKr cells (0.8\% of channels) and temporarily
masked LKr cells is used. Particle identification, trigger and
readout efficiencies are measured directly from data.

Charged particle tracks are reconstructed from hits and drift times
in the spectrometer. Track momenta are evaluated using a detailed
magnetic field map. Clusters of energy deposition in the LKr are
found by looking at the maxima in the digitized pulses from
individual cells in both space and time, and accumulating the energy
within a radius of 11~cm. Shower energies are corrected for energy
outside the cluster boundary, energy lost in non-working cells, and
cluster energy sharing.

Due to the topological similarity of $K_{e2}$ and $K_{\mu 2}$
decays, a large part of the selection is common for the two modes:
(1) exactly one reconstructed particle of positive electric charge
geometrically consistent with originating from a kaon decay is
required; (2) extrapolated track impact points in the DCHs, HOD and
LKr must be within their geometrical acceptances; (3) track momentum
must be in the range (13; 65)~GeV/$c$, where the lower limit assures
the efficiency of the $E_\mathrm{LKr}>10$~GeV trigger condition; (4)
no LKr energy deposition clusters with energy
$E>E_\mathrm{veto}=2$~GeV and in time with the track are allowed
unless they are consistent with being produced by the track via
direct energy deposition or bremsstrahlung; (5) the reconstructed
decay vertex longitudinal position must be within the nominal decay
volume; (6) distance between the charged track and the nominal kaon
beam axis must be below 3.5~cm.

The following two principal criteria are used to distinguish
$K_{e2}$ from $K_{\mu 2}$ decays. Kinematic identification is based
on constraining the reconstructed squared missing mass in positron
(muon) hypothesis $-M_1^2 < M_{\mathrm{miss}}^2(\ell) = (P_K -
P_\ell)^2 < M_2^2$, where $P_K$ and $P_\ell$ are the four-momenta of
the kaon (defined as the average one monitored with
$K^+\to\pi^+\pi^+\pi^-$ decays) and the lepton (under the $e^+$ or
$\mu^+$ mass hypothesis). The limits $M_1^2$ and $M_2^2$ have been
optimized taking into account the resolution and backgrounds, and
vary among lepton momentum bins in the ranges (0.013; 0.016) and
(0.010; 0.014) (GeV/$c^2$)$^2$, respectively. Lepton identification
is based on the ratio $E/p$ of track energy deposition in the LKr to
its momentum measured by the spectrometer. Tracks with
$(E/p)_\mathrm{min}<E/p<1.1$, where $(E/p)_\mathrm{min}=0.95$ for
$p>25$~GeV/$c$ and $(E/p)_\mathrm{min}=0.9$ otherwise, are
identified as positrons. Tracks with $E/p<0.85$ are identified as
muons.

\section{Backgrounds}

\begin{figure}[tb]
\begin{center}
\resizebox{0.45\textwidth}{!}{\includegraphics{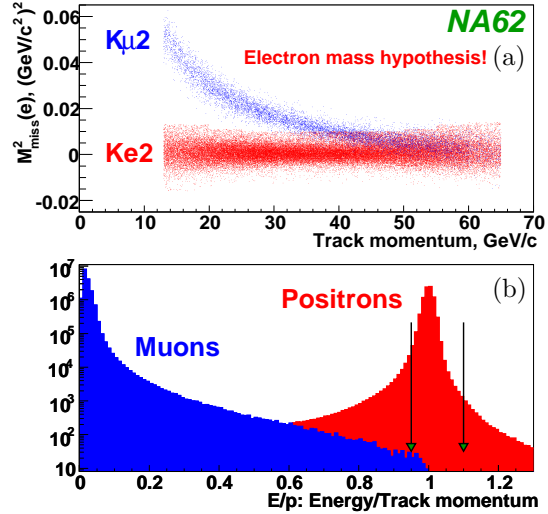}}
\put(-22,170){(a)} \put(-22,82){(b)} \vspace{-10mm} \caption{(a)
$M^2_{\rm miss}(e)$ vs lepton momentum for reconstructed $K_{e2}$
and $K_{\mu2}$ decays; (b) $E/p$ spectra of positrons and muons.}
\end{center}\vspace{-10mm}
\end{figure}

Kinematic separation of $K_{e2}$ from $K_{\mu 2}$ decays is
achievable at low lepton momentum only ($p<30$~GeV/$c$), as shown in
Fig.~1a. At high lepton momentum, the $K_{\mu2}$ decay with the muon
mis-identified as positron ($E/p>0.95$, as shown in Fig.~1b) due to
`catastrophic' bremsstrahlung in or in front of the LKr is the
largest background source. In order to measure the
mis-identification probability $P_{\mu e}$, a muon sample free from
the typical $\sim10^{-4}$ positron contamination due to $\mu\to e$
decays has been collected: a $9.2X_0$ thick lead (Pb) wall covering
$\sim 20\%$ of the geometric acceptance was installed in front of
the LKr during a period of data taking. The positron component in a
sample of muon candidates consistent with originating from
$K_{\mu2}$ decays, traversing the Pb wall, with $p>30$~GeV/$c$ and
$E/p>0.95$ is suppressed to a negligible level ($\sim 10^{-8}$) by
energy losses in the Pb wall.

However, the muon passage through the Pb wall affects the measured
$P_{\mu e}^\mathrm{Pb}$ via two principal effects: 1) ionization
energy loss in Pb decreases $P_{\mu e}$ and dominates at low
momentum; 2) bremsstrahlung in Pb increases $P_{\mu e}$ and
dominates at high momentum. To evaluate the corresponding correction
factor $f_\mathrm{Pb}=P_{\mu e}/P_{\mu e}^\mathrm{Pb}$, a dedicated
Geant4 based~\cite{geant4} MC simulation of muon propagation
downstream the spectrometer involving all electromagnetic processes,
including muon bremsstrahlung~\cite{ke97}, has been developed. The
relative systematic uncertainties on $P_{\mu e}$ and $P_{\mu
e}^{\mathrm{Pb}}$ obtained by simulation are estimated to be $10\%$,
mainly due to the simulation of cluster geometry and calibration
(measured and simulated $P_{\mu e}^{\mathrm{Pb}}$ are shown in
Fig.~2). However, the error affecting their ratio is significantly
smaller ($\delta f_\mathrm{Pb}/f_\mathrm{Pb}=2\%$) due to partial
cancellation of uncertainties.

\begin{figure}[tb]
\begin{center}
\resizebox{0.45\textwidth}{!}{\includegraphics{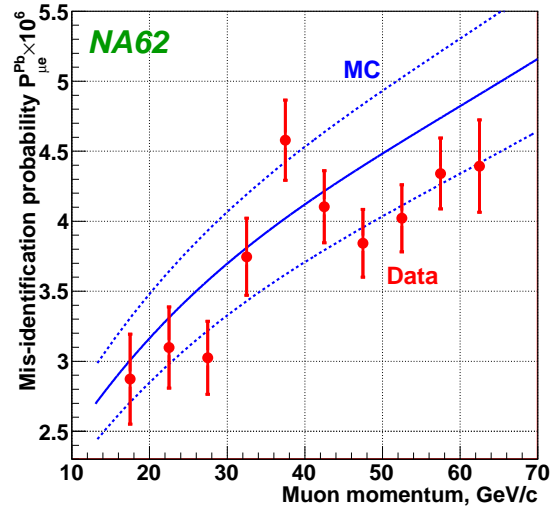}}
\vspace{-14mm}\end{center} \caption{Muon mis-identification
probability with the Pb wall installed $P_{\mu e}^\mathrm{Pb}$ for
$(E/p)_\mathrm{min}=0.95$ vs momentum: measurement (markers with
error bars); simulation with its uncertainty (solid and dashed
lines).}\vspace{-7mm} \label{fig:pmue}
\end{figure}

The $K_{\mu 2}$ background contamination has been computed to be
$(6.10\pm0.22)\%$ using the measured $P_{\mu e}^\mathrm{Pb}$
corrected by the simulated $f_\mathrm{Pb}$, and correcting for the
correlation between the reconstructed $M_\mathrm{miss}^2(e)$ and
$E/p$. The uncertainty comes from the limited size of the data
sample used to measure $P_{\mu e}^\mathrm{Pb}$ (0.16\%), the
uncertainty $\delta f_\mathrm{Pb}$ (0.12\%), and  model-dependence
of the $M_\mathrm{miss}^2(e)$ vs $E/p$ correlation (0.08\%).

The $K_{\mu2}$ decay also contributes to background via the $\mu\to
e$ decay in flight. Energetic forward daughter positrons compatible
with $K_{e2}$ kinematics and topology are suppressed by muon
polarization effects~\cite{mi50}. Radiative corrections to the muon
decay~\cite{ar02} lead to a further $\sim10\%$ relative background
suppression. The background contamination has been estimated to be
$(0.27\pm0.04)\%$.

The structure-dependent (SD) $K^+\to e^+\nu\gamma$
process~\cite{bi93}, not suppressed by angular momentum conservation
(more specifically, its $\mathrm{SD}^+$ component corresponding to
positive photon helicity), represents a significant background
source. A recent measurement of the $K^+\to
e^+\nu\gamma~(\mathrm{SD}^+)$ differential decay rate~\cite{am09}
has been used to evaluate the background contamination to be
$(1.15\pm0.17)\%$. The dominant uncertainty comes from the
uncertainty on the rate of $K^+\to e^+\nu\gamma~(\mathrm{SD^+})$
decay, which has been increased by a factor of 3 with respect to
that reported in~\cite{am09}, as suggested by a stability check of
$R_K$ with respect to a variation of the $E_\mathrm{veto}$ limit.

The beam halo background is induced by halo muons undergoing $\mu\to
e$ decays in the vacuum tank, or being mis-identified as positrons.
It has been measured directly by reconstructing the $K^+_{e2}$
candidates from a $K^-$ data sample collected with the $K^+$ beam
(but not its halo) blocked, and a special data sample collected with
both beams blocked. The control sample is normalized to the data in
the region $-0.3<M_\mathrm{miss}^2(\mu)<-0.1$ $(\mathrm{GeV}/c^2)^2$
populated predominantly by beam halo events. The `cross-talk'
probability to reconstruct a $K_{e2}^+$ candidate due to a $K^-$
decay with $e^+$ emission ($K^-\to\pi^0_D\ell^-\nu$,
$K^-\to\pi^-\pi^0_D$, $K^-\to\ell^-\nu e^+e^-$, where $\pi^0_D$
denotes the $\pi^0$ Dalitz decay $\pi^0\to\gamma e^+e^-$) is at the
level of $\sim 10^{-4}$ and is taken into account. The halo
background contamination has been estimated to be $(1.14\pm0.06)\%$,
where the uncertainty comes from the limited size of the control
sample and the uncertainty of its normalization. The beam halo is
the only significant background source in the $K_{\mu2}$ sample,
measured to be $(0.38\pm0.01)\%$ with the same technique.

The number of $K_{\ell 2}$ candidates is $N(K_{e2}) = 59,963$ (about
four times the statistics collected by KLOE) and $N(K_{\mu2}) =
1.803\times 10^7$. The $M_{\rm miss}^2(e)$ distributions of data
events and backgrounds are presented in Fig.~3; backgrounds in the
$K_{e2}$ sample integrated over lepton momentum are summarized in
Table 1.

\begin{figure}[tb]
\begin{center}
\resizebox{0.45\textwidth}{!}{\includegraphics{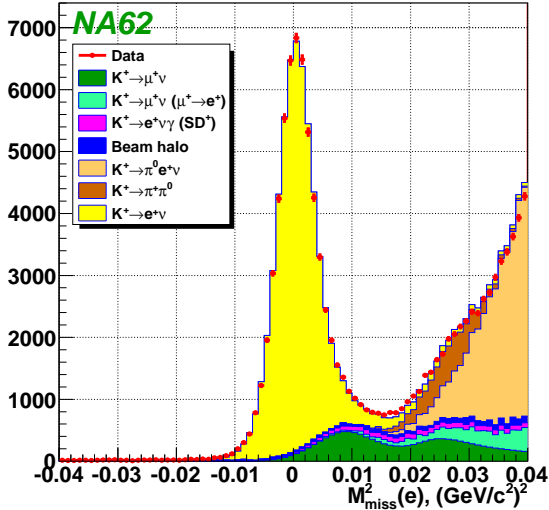}}
\vspace{-12mm}\end{center} \caption{Reconstructed squared missing
mass $M_{\mathrm{miss}}^2(e)$ distribution of the $K_{e2}$
candidates compared with the sum of normalised MC signal and
background components.}\vspace{-7mm}\label{fig:mm2e}
\end{figure}

\begin{table}[tb]
\caption{Summary of backgrounds in the $K_{e2}$ sample.}
\begin{tabular}{lc}
\hline Source & $N_B/N(K_{e2})$\\
\hline
$K_{\mu2}$                     & $(6.10\pm0.22)\%$\\
$K_{\mu2}~(\mu\to e)$          & $(0.27\pm0.04)\%$\\
$K_{e2\gamma}~(\mathrm{SD}^+)$ & $(1.15\pm0.17)\%$\\
Beam halo                      & $(1.14\pm0.06)\%$\\
$K_{e3}$                       & $(0.06\pm0.01)\%$\\
$K_{2\pi}$                     & $(0.06\pm0.01)\%$\\
\hline
Total background & $(8.78\pm0.29)\%$\\
\hline \vspace{-10mm}
\end{tabular}
\end{table}

\section{Systematic uncertainties}

The ratio of geometric acceptances $A(K_{\mu2})/A(K_{e2})$ in each
lepton momentum bin has been evaluated with a MC simulation. The
radiative $K^+\to e^+\nu\gamma$ (IB) process is simulated
following~\cite{bi93} with higher order corrections according
to~\cite{we65,ga06}. Lepton tracking inefficiency due to
interactions with the spectrometer material is included into the
acceptance correction, and its simulation has been validated with
the data. The main sources of systematic uncertainty of the
acceptance correction are the limited knowledge of beam profile and
divergence, accidental activity, and the simulation of soft
radiative photons. A separate uncertainty has been assigned due to
the limited precision of the DCH alignment.

A sample of $\sim 4\times 10^7$ positrons selected kinematically
from $K^+ \to \pi^0 e^+ \nu$ decays collected concurrently with the
main $K_{\ell 2}$ data set is used to calibrate the energy response
of each LKr cell, and to study $f_e$ with respect to local position
and time stability (in the kinematically limited momentum range
$p<50~\mathrm{GeV}/c$). A sample of electrons and positrons from the
$4\times10^{6}$ $K_L \to \pi^\pm e^\mp \nu$ decays collected during
a special short (15h) run with a broad momentum band $K^0_L$ beam
allows the determination of $f_e$ in the whole analysis momentum
range. The measurements of $f_e$ have been performed in bins of
lepton momentum; separate measurements have been performed for
several identified groups of LKr cells with higher local
inefficiencies. The inefficiency averaged over the $K_{e2}$ sample
is $1-f_e = (0.73\pm0.05)\%$, where the uncertainty takes into
account the statistical precision and the small differences between
$K^+$ and $K^0_L$ measurements.

The efficiency of the $Q_1$ trigger condition has been measured
using $K_{\mu2}$ events triggered with a special control LKr signal:
integrated over the $K_{\mu2}$ sample, it is $(1.4\pm0.1)\%$. Owing
to its geometric uniformity, and the similarity of the $K_{e2}$ and
$K_{\mu2}$ distributions over the HOD plane, it mostly cancels
between the $K_{e2}$ and $K_{\mu2}$ samples, and the residual
systematic bias on $R_K$ is negligible. The inefficiency of the
1-track trigger for $K_{\ell 2}$ modes is negligible. The trigger
efficiency correction $\epsilon(K_{\mu2})/\epsilon(K_{e2})$ is
determined by the efficiency $\epsilon(E_\mathrm{LKr})$ of the LKr
energy deposit trigger signal $E_\mathrm{LKr}>10$~GeV, which has
been measured to be $1-\epsilon(E_\mathrm{LKr}) = (0.41\pm0.05)\%$
in the first lepton momentum bin of (13; 20) GeV/$c$, and to be
negligible in the other momentum bins. The corresponding uncertainty
on $R_K$ is negligible.

\begin{figure}[tb]
\begin{center}
\resizebox{0.45\textwidth}{!}{\includegraphics{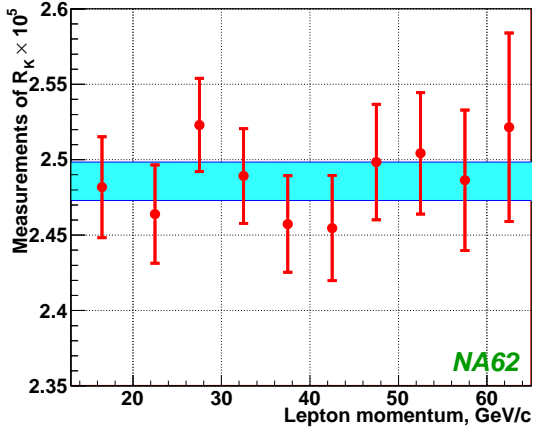}}
\vspace{-12mm}\end{center} \caption{Measurements of $R_K$ in lepton
momentum bins, and the averaged $R_K$ (indicated by a band). The
uncertainties in momentum bins are partially correlated, and include
statistical and systematic contributions.}\vspace{-7mm}
\end{figure}

\begin{table}[tb]
\caption{Summary of the uncertainties on $R_K$.}
\begin{tabular}{lc}
\hline Source & $\delta R_K\times 10^5$\\
\hline Statistical         & 0.011\\
$K_{\mu2}$ background      & 0.005\\
$K^+\to e^+\nu\gamma~(\textrm{SD}^+)$ background & 0.004\\
Beam halo background       & 0.001\\
Acceptance correction      & 0.002\\
Spectrometer alignment     & 0.001\\
Positron identification    & 0.001\\
1-track trigger efficiency & 0.002\\
\hline
\end{tabular}
\vspace{-7mm}
\end{table}

Energetic photons not reconstructed in the LKr may initiate showers
by interacting in the DCH or beam pipe material, which causes the
DCH hit multiplicities to exceed the limits allowed by the 1-track
trigger condition. This suppresses the $K^+\to
e^+\nu\gamma~(\mathrm{SD}^+)$ background by about 10\% relative
(varying over the positron momentum). Evaluation of the 1-track
inefficiency for $K^+\to e^+\nu\gamma~(\mathrm{SD}^+)$ partially
relies on simulation; its uncertainty has been propagated into
$R_K$.

The global LKr readout inefficiency has been measured using an
independent readout system to be $1-f_\mathrm{LKr}=(0.20\pm0.03)\%$
and stable in time.

\section{Result and conclusions}

The independent measurements of $R_K$ in the 10 lepton momentum bins
and the average over the bins are displayed in Fig.~4. Extensive
stability checks in bins of kinematic variables, against variation
of selection criteria and analysis procedures have been performed.
The uncertainties of the combined result are summarized in Table~2.
The result is
\begin{eqnarray*}
R_K&=&(2.486\pm 0.011_{\mathrm{stat.}} \pm
0.008_{\mathrm{syst.}})\times 10^{-5}\\
&=&(2.486\pm0.013)\times 10^{-5}.
\end{eqnarray*}
This is the most precise measurement to date; it is consistent with
the SM expectation.


\end{document}